# Persistent metal-insulator transition at the surface of an oxygen-deficient, epitaxial manganite film

Paul C. Snijders,[1,†] Min Gao,[1,2,†,‡] Hangwen Guo,[1] Guixin Cao,[3] Wolter Siemons,[1] Hongjun Gao,[2] Thomas Z. Ward,[1] Jian Shen,[4] and Zheng Gai [1,3*]

1 Materials Science and Technology Division, Oak Ridge National Laboratory, Oak Ridge, Tennessee 37831, USA,

2 Institute of Physics, Chinese Academy of Sciences, Beijing, 100190, China

3 Center for Nanophase Materials Sciences, Oak Ridge National Laboratory, Oak Ridge, Tennessee 37831, USA

4 State Key Laboratory of Surface Physics and Department of Physics, Fudan University, Shanghai 200433, China

*gaiz@ornl.gov

†These authors contributed equally to this work

‡ Present address: State Key Laboratory of Electronic Thin Films and Integrated Devices, University of Electronic Science and Technology of China , Chengdu,  Sichuan 610054, P. R. China,

Abstract

The oxygen stoichiometry has a large influence on the physical and chemical properties of complex oxides. Most of the functionality in *e.g.* catalysis and electrochemistry depends in particular on control of the oxygen stoichiometry. In order to understand the fundamental properties of intrinsic surfaces of oxygen-deficient complex oxides, we report on *in situ* temperature dependent scanning tunnelling spectroscopy experiments on pristine oxygen deficient, epitaxial manganite films. Although these films are insulating in subsequent *ex situ* in-plane electronic transport experiments at all temperatures, *in situ* scanning tunnelling spectroscopic data reveal that the surface of these films exhibits a metal-insulator transition (MIT) at 120 K, coincident with the onset of ferromagnetic ordering of small clusters in the bulk of the oxygen-deficient film. The surprising proximity of the surface MIT transition temperature of nonstoichiometric films with that of the fully oxygenated bulk suggests that the electronic properties in the surface region are not significantly affected by oxygen deficiency in the bulk. This carries important implications for the understanding and functional design of complex oxides and their interfaces with specific electronic properties for catalysis, oxide electronics and electrochemistry.

In catalysis and electrochemistry the surface properties of catalyst and electrode materials dominate their functional performance. For oxide materials this is often associated with the bulk oxygen stoichiometry, with oxygen vacancies frequently improving the catalytic and electrochemical performance [1, 2]. This contrasts with the spectacular physical properties with promising functionality displayed by complex oxides such as perovskites [3-6] but that are generally strongly reduced or even eliminated by the introduction of oxygen vacancies [7]. Moreover, in nanoscale oxide heterostructures the possible large chemical and electrostatic potential gradients, and the high surface/volume ratios may drastically affect the properties of the material[8]. With respect to the chemically important surface properties, it is currently not clear whether an oxygen deficiency in the bulk of a three-dimensional (3D) perovskite affects these surface properties in concert with those of the bulk, or whether this is significantly different due to *eg*. altered oxygen vacancy formation energies near the surface due to different local ionic coordination or even a different local stoichiometry of the unit cells at the surface and interface.

Complicating our understanding of the electronic properties of catalytically relevant complex oxide surfaces is the fact that at their surfaces and interfaces the broken translational symmetry fundamentally changes the properties[9, 10]. While this has been shown to produce remarkable physics at interfaces [11-13], at surfaces of three-dimensional (3D) perovskites the results are qualitatively different: many fully oxygenated 3D perovskite surfaces and interfaces exhibit a so-called nanoscale "dead layer" where the magnetic order and metal-insulator transitions that from the bulk are strongly reduced or even absent [14-19]. In first order, this can be explained by the decreased coordination at the surface that decreases the electronic band width, increases electronic correlations, and creates insulating phases. This picture is further obfuscated by the structural relaxations at the surface and charge transfer to or from the bulk [20, 21] which affect phonon frequencies, and the symmetry, dispersion, and filling of electronic bands, and (double) exchange interaction [14, 20, 21]. Finally, the surface chemistry and surface defect concentration will affect the physical properties through their impact on the oxidation state of the transition metal cations. These changes thus not only influence the electronic and magnetic contact properties in heterostructure devices [16, 17], but most importantly will affect the surface chemical or catalytic characteristics as well.

Moreover, the absence of good cleaving properties in 3D perovskites forces one to have *in situ* experimental surface analysis techniques in order to measure the properties of pristine surfaces and prevent further complications by contamination from exposure to the ambient atmosphere [22]. To our knowledge the basic electronic properties of unexposed, pristine surfaces of catalytically and electrochemically important oxygen-deficient 3D perovskite materials have not been studied so far; a clean starting point for understanding catalysis and electrochemistry using perovskite oxides is therefore lacking.

Here we have measured the temperature-dependent surface electronic properties of pristine oxygen deficient $(La_{1-x}Pr_x)_{5/8}Ca_{3/8}MnO_{3-\delta}$ (x=0.3) (LPCMO) thin films using temperature dependent scanning tunneling spectroscopy (STS). These films were grown and analyzed *in situ*, *i.e.* without exposing their surfaces to ambient conditions. The surface exhibits an unexpected temperature dependent metal-insulator transition (MIT) even though the bulk of the films remains insulating at all temperatures, suggesting that the intrinsic electronic properties in the surface region are not significantly affected by an oxygen deficiency in the bulk.

Results and Discussions

The films were grown by pulsed laser deposition (PLD) on 0.05 wt % Nb-doped $SrTiO_3$(001) substrates. The choice for STO substrates is motivated by the resulting tensile strain in the LPCMO films [23] that facilitates oxygen vacancy formation [24]. Prior to PLD growth, the substrates were chemically etched in buffered HF and then annealed for 3 hours at 950ºC in an oxygen atmosphere. During growth, the substrate was kept at about 825ºC in a flowing oxygen (10% ozone) environment under a pressure $\sim 1\times 10^{-3}$ Torr. After growth, the samples were slowly cooled down to 600 ºC at the same oxygen pressure used during growth, after which they were quickly cooled down to room temperature in vacuum, introducing oxygen vacancies in the bulk of the film. Note that oxygen vacancy concentrations in oxide thin films tend to decrease significantly near the surface [25]. Reflection high energy electron diffraction (RHEED) was used to detect the quality and thickness of the films during growth, revealing layer-by-layer intensity oscillations and a 1x1 pattern at all thicknesses. The absence of a reconstruction indicates that indeed no oxygen-vacancy induced reconstructions exist at the surface. Atomically ordered single crystalline films have been produced routinely using similar procedures in our lab [26, 27]. After growth the sample was transferred to a connecting variable temperature scanning tunneling microscopy (STM) chamber without breaking ultra-high vacuum. All STM images and tunneling spectra were obtained using mechanically cut Pt-Ir tips. The crystalline quality and lattice parameters of the bulk of the films were measured ex situ by X-ray diffraction (XRD). Resistivity vs. temperature curves were measured by a physical property measurement system (PPMS) under magnetic fields up to 9 T with a constant current of 100 nA in a 4-probe configuration. Magnetization data were taken using a Quantum Design magnetic properties measurement system (MPMS).

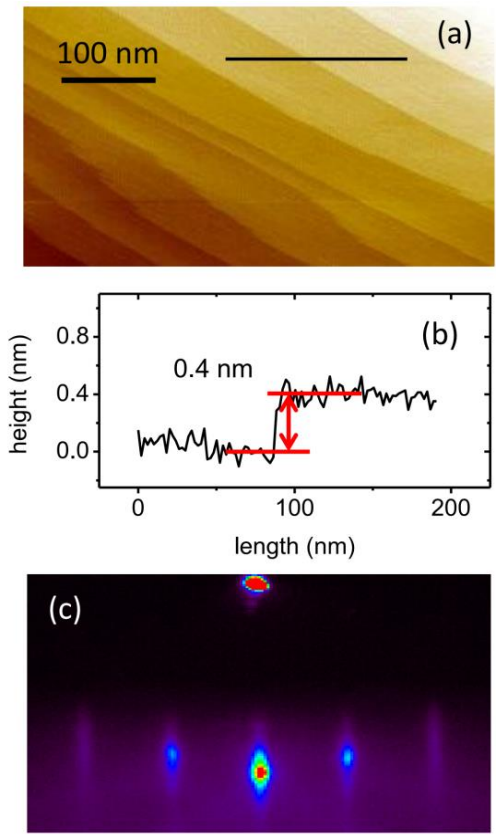

*Figure 1. (a) STM image of a 100 nm thick film ($V_s$=-0.5 V, I=30 pA). (b) Height profile along the line in (a) showing steps of unit cell height within the precision of our measurement. (c) RHEED image from the grown film.*

Figure 1(a) shows the typical morphology of a 100 nm thick LPCMO(001) film. The morphology reveals a clean layer-by-layer growth mode. The height profile along the line in Figure 1(a) is plotted in Figure 1(b). The step height of all steps at the surface is ~0.4 nm, equal to the lattice constant of LPCMO, showing that only one kind of termination exists at the surface, in agreement with the persistent 1x1 structure evidenced by our *in situ* RHEED pattern, Figure 1(c). Following previous work [28] we conclude our films are terminated at the Mn-O plane as it was grown on $TiO_2$ terminated STO. No indication of segregated oxide clusters (e.g. CaO) was observed in either XRD or STM images.

Figure 2 presents a reciprocal space map (a) collected around the (103) reflection of the sample, and a θ-2θ scan (b), confirming the single crystalline nature of the grown film. We emphasize here that the XRD data pertain to the bulk of the film as the system sensitivity for measuring the lattice parameters at the surface alone is too low. The data indicate that the film, even at 100 nm thickness, is not relaxing significantly with increasing distance from the substrate but is still coherently strained to the in-plane lattice constant of 3.905 Å imposed by the epitaxial growth on STO. The oxygen distribution appears homogeneous and random as no structural phase separation (e.g. into a Brownmillerite structure) is detected. The c-parameter in the film is contracted as compared to bulk LPCMO of this doping [29], as expected for a tensile-strained film grown on STO. However, the calculated Poisson ratio for the biaxially strained film $(a,b,c)_{film}$ = (3.90,3.90,3.8255) Å using $(a,b,c)_{bulk}$ = (3.848,3.849,3.840) Å for freestanding bulk LPCMO [29] is 0.123. This is most likely due to the presence of a large concentration of oxygen vacancies [30], which may stabilize the epitaxially strained lattice matching up to large thicknesses. Comparing our data with the relative change in unit cell volume per relative change in oxygen deficiency in related $La_{0.65}Ca_{0.35}MnO_3$ from the inset in Figure 3 of Ref. [31], we estimate our oxygen vacancy concentration to be 12%, i.e. δ=0.36 in LPCMO3-δ based on our volume change of 2.25%. This number should be regarded with caution though, as the biaxial tensile strain of the epitaxial film tends to decrease the c lattice parameter. The tensile strain in the film and the large oxygen vacancy concentration are likely mutually stabilizing factors [24] and their respective roles here are intertwined. This makes an accurate determination of δ in the bulk of the film difficult.

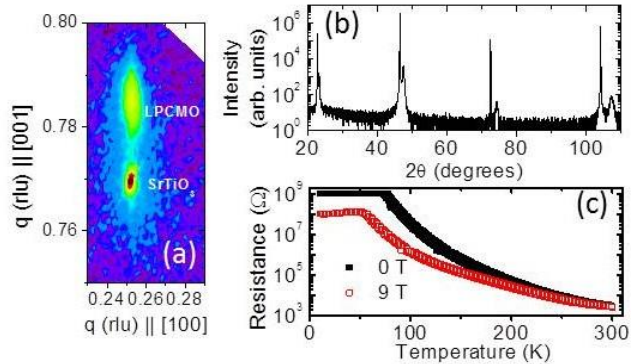

Figure 2. *(a) Reciprocal space map of a 100 nm thick epitaxial LPCMO film collected around the (103) reflection, showing good epitaxial lattice matching. (b) θ-2θ wide scan showing the absence of oxygen vacancy induced crystal structures. (c) The temperature dependence of the resistance measured using the PPMS at 0 and 9 T for the film from which the spectroscopic data are presented in Figure 3. Our measurement capability is limited to $1\times10^9$ Ohm, which is reached at about 78 K for the black curve.*

Like other colossal magnetoresistive manganites[5, 6], bulk stoichiometric LPCMO exhibits a paramagnetic to ferromagnetic transition concomitant with a MIT at a transition temperature of 120 K [32-34]. Both tensile strain [23, 35] and the introduction of oxygen vacancies shift the MIT to lower temperatures [7], the former through a stabilization of the charge ordered (Jahn-Teller distorted) state at the expense of the ferromagnetic metallic state [35], and the latter through a direct structural disruption of the ferromagnetic double exchange interaction, essentially destroying the domain structure of LPCMO [7]. In Figure 2(c) we present an in-plane resistance measurement as a function of temperature that was obtained *ex situ* using a PPMS after the *in situ* STS experiments were completed, see below. Our epitaxial, single crystalline LPCMO films grown under tensile strain on STO with an intentionally large oxygen deficiency are insulating at all temperatures down to 10 K, and have an exceptionally high resistance beyond our measurement capabilities below 78 K. In a high magnetic field of 9 T a MIT reminiscent of but at a lower temperature than that of stoichiometric bulk reappears (Figure 2(c)).

Figure 3(a) shows temperature dependent tunneling spectroscopy I(V) data obtained *in situ* from the surface of the pristine LPCMO film at temperatures between 75 K and 250 K during warming. We ensured that our spectra are representative of the average surface properties by averaging many hundreds of I(V) spectra in a sampling area larger than the micron length scale associated with electronic phase separation in fully oxygenated bulk LPCMO [33]: individual I(V) spectra were recorded using grids of spectroscopy locations with individual curves separated between 1 to 5 nm. By recording multiple adjacent grids, total areas of up to 10 by 10 microns were covered. Analyzing the individual curves, we find no indication of electronic phase separation being present. Instead the surface electronic properties appear to be homogeneous, consistent with the expected effect of oxygen vacancies destroying long-range electronic phase separation in the bulk [7]. From Figure 3(a) it is clear that the tunneling current at higher bias voltage has a minimum value between 100 K and 120 K. The three I(V) spectra presented in Figure 3(b) confirm this observation, with the dashed curve recorded at 106 K showing a significantly lower current at high tunneling biases. A similar temperature dependence of the I(V) curves near the metal-insulator transition temperature ($T_{MIT}$) was also observed by others on related $La_{0.7}Ca_{0.3}MnO_3$ films [36]. Following the reasoning in this work we conclude this high-bias minimum is a signature of a metal-insulator phase transition in the surface DOS of our oxygen deficient LPCMO films.

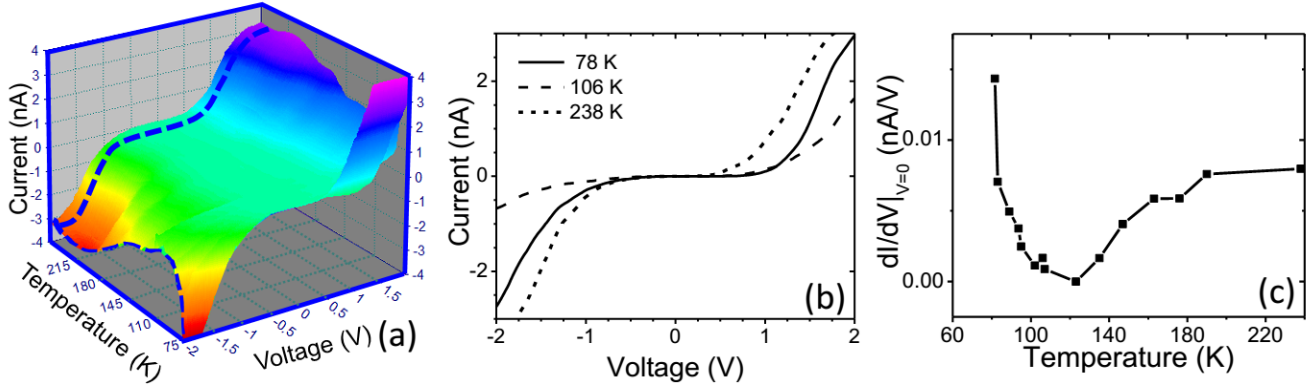

Figure 3. *(a) 3D representation of the I(V) tunneling spectra measured as a function of temperature on a 100 nm thick LPCMO film. One I(V) spectrum and a guide for the eye for the current as a function of temperature at -2V are plotted with a dashed blue line for clarity. The color presents the magnitude of tunneling current, and corresponds to the scale on the current axis. All of the curves were measured using $V_s$=-0.5 V, I=30 pA. (b) Three selected I(V) curves above (238K), near (106K), and below (78K) the MIT. (c) The temperature dependence of the zero bias conductance $d(I)/d(V)|_{V=0}$ extracted from the data in (a).*

To confirm the presence of a MIT, we focus on the temperature dependence of $d(I)/d(V)|_{V=0}$ as the large dynamic range in the I(V) spectra prevents a proper visualization of the low bias characteristics Moreover, this zero bias conductance (ZBC) in tunneling spectroscopy reflects the magnitude of the DOS at the Fermi level that governs transport properties in a uniform material. Figure 3(c) is the ZBC of LPCMO film extracted from Figure 3(a). The ZBC displays a strong temperature dependence, decreasing below 190 K, reaching a minimum value around 120 K [36], in close agreement with $T_{MIT}$ in Figure 3(a), and increasing rapidly with decreasing temperature below 120 K. The increase in ZBC below $T_{MIT}$ can be attributed to changes in the DOS due to an increase in delocalized carrier density or a decrease in polaron density in the metallic phase as proposed in Refs. [36-38]. The increase in ZBC above the $T_{MIT}$ is consistent with that observed in $La_{0.7}Ca_{0.3}MnO_3$ [36], where it was attributed to thermal broadening of the spectroscopic data in the presence of a small polaronic gap, see Refs. [36, 38] and the discussion below.

Clearly a zero-field temperature dependent MIT is present at the surface of oxygen-deficient, insulating, 3D perovskite LPCMO films with a Mn-O termination. Its transition temperature is unexpectedly close to that of fully oxygenated bulk LPCMO [33]. This observation indicates that the broken symmetry due to the presence of the (pristine) surface does not necessarily result in a nanoscale dead layer, in contrast with the general notion that insulating and/or non-magnetic surface dead layers are ubiquitous in complex oxides [15, 19]. More importantly, these findings demonstrate that the intrinsic electronic properties of the surface of severely oxygen deficient complex oxides can be surprisingly similar to those of fully oxygenated bulk. In this respect, it is interesting to note that we have shown earlier in Ref. [26] that the pristine Mn-O terminated surface of stoichiometric $La_{5/8}Ca_{3/8}MnO_3(001)$ is metallic at room temperature, i.e. above than the ferromagnetic metal to paramagnetic insulator transition temperature of the bulk ($T_C$=260 K [26]), presumably caused by the broken symmetry at the surface. Moreover, in Ref. [39], an ambient-exposed LPCMO surface was observed to exhibit a gap in tunneling spectroscopy data even below the bulk MIT transition temperature. These results confirm our premise that it is imperative to study complex oxide surfaces without exposure to the ambient in order to achieve a basic understanding of their intrinsic properties [40]. Indeed surfaces exposed to the ambient atmosphere almost always appear to exhibit smooth, parabolic or V-shaped scanning tunneling spectroscopy dI/dV curves [36-39, 41, 42], even in energy ranges where narrow Mn 3d states should produce pronounced peaks in the spectroscopic data. Often a (soft) gap is observed even in the metallic phase below $T_{MIT}$ [37-39, 43], which is generally interpreted in terms of a soft (pseudo-) gap due to small Jahn-Teller (JT) polarons. However, some of these results were recognized to be ambiguous because a contamination-induced fully insulating surface layer could allow the detection of the transition in the bulk by acting as an addition to the vacuum tunneling barrier in STS experiments [37, 38], and the influence of surface contamination could not be ruled out [39].

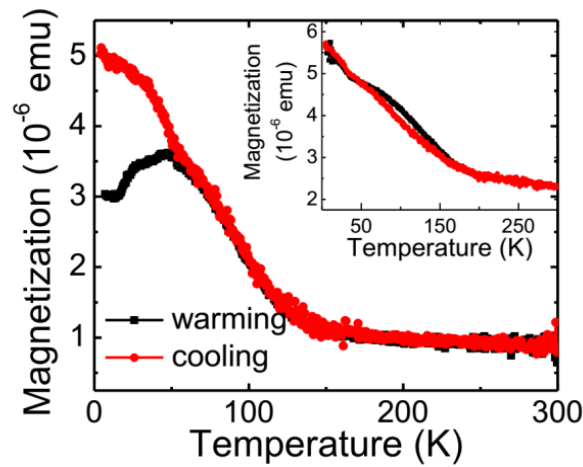

Figure 4. *(a) Warming (after zero field cooling) and cooling magnetization curves of the as-grown LPCMO film in a 1000 Oe field. The inset shows the corresponding M(T) curves of the same sample after annealing in 1 bar flowing oxygen for 8 hours.*

In order to investigate whether the observed surface MIT on oxygen deficient LPCMO is coupled to the 3D bulk magnetic degrees of freedom, we have performed *ex situ* temperature dependent magnetization measurements on our films as shown in Figure 4. They reveal a sudden appearance of a ferromagnetic phase around $T_C$=120 K, i.e. at the same temperature as the surface $T_{MIT}$. The absence of thermal hysteresis in the cooling and warming magnetization curves in Figure 4 implies that large scale domain growth is inhibited in our oxygen deficient thin films and the observed magnetization originates in small, cluster-like regions [44]. Consistently, an oxygen anneal of the sample completely recovers the thermal hysteresis in the temperature dependent magnetization expected in the presence of large domains (see the inset in Figure 4), increases the $T_{MIT}$ towards that of fully oxygenated LPCMO films on STO [23], and changes the lattice c parameter back to c=3.795 Å giving an expected Poisson ratio of 0.304. Hence we conclude that our films differ from "good" LPCMO films only by their oxygen deficiency. The onset temperature and magnitude of the magnetization remains comparable for our severely oxygen deficient and fully oxygenated films, see the inset in Figure 4. This suggests that a competition similar to that from which colossal magnetoresistance in stoichiometric manganites originates, is locally still present in the bulk of our strained, oxygen deficient films but that long range ordered domain formation is inhibited.

However, despite the conjecture above, the nature of the metallic and insulating phases we observe here remains tentative; a possible electron doping effect due to the oxygen vacancies in the bulk [45], a potentially decreased oxygen vacancy concentration near the surface [25], and an altered balance of competing interactions due to the broken symmetry at the surface [9, 10, 15] all may affect the nature of the observed metallic and insulating states. The absence of surface magnetic and structural probes in our vacuum system, make it impossible to assess (near-) surface magnetic and structural properties without exposing the surface to the ambient. We can nonetheless extract a few clues as a starting point for further studies into the nature of the observed metallic and insulating surface phases and the origins of the observed surface MIT.

The metallic phase observed here at low temperatures could well be related to the metallicity observed on the Mn-O terminated $La_{5/8}Ca_{3/8}MnO_3$(001) surface [26]; while some of the La is replaced by Pr in the current study, promoting a charge ordered (JT) insulating state [6], the hole doping level is the same in these materials. The absence of translational symmetry perpendicular to the surface plane and the termination of the surface in the basal Mn-O plane instead of a full $MnO_6$ octahedron, create a square pyramidal $C_{4v}$ symmetry of the surface unit cell. This forces a preferential orbital occupation of the Mn $3d_{3z2-r2}$ orbitals near the surface [46, 47], resulting in a larger overlap with the neighboring oxygen $2p$ orbitals and a metallic conductivity channel perpendicular to the surface [26]. As the effect of the symmetry breaking is not strictly limited to the very surface layer itself, the preferential orbital occupation can extend a number of unit cells below the surface [15, 47]. This may even be further enhanced by the compressive stress along the c-axis that is associated with the epitaxially imposed in-plane tensile strain which increases the bandwidth for momenta perpendicular to the plane of the film. Also the presence of oxygen vacancies could enhance the coupling to the bulk, depending on their energetically preferred position (apical or basal) in near-surface unit cells. Note that the strain dependence of the relative basal and apical oxygen vacancy formation energies depends sensitively on the balance between crystal field and electrostatic effects [48], and may thus vary from one perovskite to another [24, 48]. In the apical position oxygen vacancies relieve compressive strain along the c-axis, and the resulting local $C_{4v}$ symmetry (rotation axis along the lattice c-axis) results in a hybridization of the Mn $4_s$ and $4_p$ orbitals with the Mn $3d_{3z2-r2}$ orbitals and enhanced Mn-Mn hopping integrals along the c-axis, while keeping the other Mn 3d orbitals relatively unaffected [49]. Note that the inherent out of plane anisotropy of these conducting channels would favor detection by STS and could possibly be undetectable in in-plane transport experiments (Figure 2(c)). Similar anisotropic behavior has been predicted in a situation of anisotropic in-plane strain [50], where it was inferred that large-scale domain percolation is not essential to obtain highly anisotropic resistivities in manganites. Instead Dong et al. [50] concluded that such anisotropy originates in anisotropic double exchange and anisotropic Jahn-Teller distortions and should be observable in a variety of manganites and other complex oxides.

Regarding the insulating phase, our data suggest that its physical origin is qualitatively different from that of dead layer insulating phases proposed in the literature. The insulating and/or non-magnetic surfaces of CMR manganite perovskites [9, 14, 15] are stable at the lowest temperatures studied and persist over the full temperature range, indicating these dead layers may be exceedingly stable ground states at these surfaces. Instead, the insulating phase we observe appears only above $T_{MIT}$, see Figure 3, suggesting an interaction-driven insulating nature, possibly similar to that in stoichiometric bulk LPCMO and $La_{1-x}Ca_xMnO_3$ (LCMO). The temperature dependence of the ZBC above $T_{MIT}$ offers us a possible clue as to its nature. Generally, when a gap opens and a material turns insulating, the DOS at the Fermi energy becomes zero. Our ZBC data do not show such behavior; instead with increasing temperature a slowly increasing finite ZBC is present, see Figure 3(b). This suggests the insulating phase is polaronic in nature, exhibiting a small pseudogap due to Jahn-Teller distortions around $Mn^{3+}$ ions above $T_{MIT}$ [36, 38]. Indeed, model calculations in Ref. [20] show that polaron formation occurs more easily on the surface than in the bulk, while the (electron-phonon) coupling strength that causes the localization (binding) of polarons associated with the MIT is enhanced at the surface. The room temperature presence of the insulating polaron phase at the pristine LPCMO surface, being absent at the LCMO surface [26] is consistent with the Pr-induced destabilization of the metallic state at the expense of the insulating state in LPCMO, naturally explaining the difference in temperature dependent properties of pristine LPCMO and LCMO surfaces. Finally, these scenarios do not need a significant vacancy incorporation at the surface to be qualitatively consistent with our observations. Indeed, the fact that the competing interactions result in a $T_{MIT}$ close to that of the fully oxygenated bulk suggests that the electronic properties at the surface of oxygen deficient manganites are not significantly affected by an oxygen deficiency in the bulk, or possibly even that the (near-) surface vacancy concentration is indeed significantly lower than in the bulk [46], consistent with the absence of vacancy induced reconstructions. These observations carry important implications for the design of complex oxide functionality in catalysis and electrochemistry.

Conclusions

In summary, we performed *in situ* temperature dependent scanning tunneling spectroscopy experiments on pristine oxygen deficient, epitaxial LPCMO films that have not been exposed to ambient conditions. The bulk of these films is insulating in *ex situ* in-plane electronic transport experiments at all temperatures. Scanning tunneling spectroscopic data reveal that the surface of these insulating films exhibits a metal-insulator transition at 120 K, coincident with the ferromagnetic ordering of small clusters in the bulk of the oxygen deficient film. The temperature dependence of the zero bias conductance above $T_{MIT}$ suggests the presence of a polaronic pseudogapped insulating phase. The broken symmetry due to the presence of the surface evidently does not induce the formation of a nanoscale dead layer on this three dimensional perovskite manganite. The surprising proximity of the surface $T_{MIT}$ of the non-stoichiometric film with that of the fully oxygenated bulk suggests that the intrinsic electronic properties in the surface region are not significantly affected by oxygen deficiency in the bulk, with important implications for the functional design of complex oxides for catalysis and electrochemistry.

Acknowledgements

We acknowledge fruitful discussions with S. Dong and C. Sen. This effort was supported by the U.S. Department of Energy (DOE), Basic Energy Sciences (BES), Materials Sciences and Engineering Division (PCS, MG, HWG, WS, TZW, ZG). A portion of this research was conducted at the Center for Nanophase Materials Sciences, which is sponsored at Oak Ridge National Laboratory by U.S. DOE, BES (GC, ZG). We also acknowledge partial funding support from the National Basic Research Program of China (973 Program) under the grant No. 2011CB921801 (JS).

References

1. M. A. Pena and J. L. G. Fierro, *Chem Rev*, 2001, **101**, 1981-2017.
2. J. Maier, *Nat Mater*, 2005, **4**, 805-815.
3. B. R. C. N. R. Rao, *Transition Metal Oxides: Structure, Properties, and Synthesis of Ceramic Oxides, 2nd Edition*, Wiley–VCH, New York and Weinheim, 1998.
4. H. Takagi and H. Y. Hwang, *Science*, 2010, **327**, 1601-1602.
5. Y. Tokura, *Rep. Prog. Phys.*, 2006, **69**, 797-851.
6. E. Dagotto, T. Hotta and A. Moreo, *Phys. Rep.-Rev. Sec. Phys. Lett.*, 2001, **344**, 1-153.
7. A. S. Ogale, S. R. Shinde, V. N. Kulkarni, J. Higgins, R. J. Choudhary, D. C. Kundaliya, T. Polleto, S. B. Ogale, R. L. Greene and T. Venkatesan, *Phys. Rev. B*, 2004, **69**, 235101.
8. S. V. Kalinin, A. Borisevich and D. Fong, *ACS Nano*, 2012, **6**, 10423-10437.
9. V. B. Nascimento, J. W. Freeland, R. Saniz, R. G. Moore, D. Mazur, H. Liu, M. H. Pan, J. Rundgren, K. E. Gray, R. A. Rosenberg, H. Zheng, J. F. Mitchell, A. J. Freeman, K. Veltruska and E. W. Plummer, *Phys. Rev. Lett.*, 2009, **103**, 227201.
10. R. G. Moore, V. B. Nascimento, J. D. Zhang, J. Rundgren, R. Jin, D. Mandrus and E. W. Plummer, *Phys. Rev. Lett.*, 2008, **100**.
11. A. Ohtomo and H. Y. Hwang, *Nature*, 2004, **427**, 423-426.

12. A. Brinkman, M. Huijben, M. Van Zalk, J. Huijben, U. Zeitler, J. C. Maan, W. G. Van der Wiel, G. Rijnders, D. H. A. Blank and H. Hilgenkamp, *Nat Mater*, 2007, **6**, 493-496.
13. N. Reyren, S. Thiel, A. D. Caviglia, L. F. Kourkoutis, G. Hammerl, C. Richter, C. W. Schneider, T. Kopp, A. S. Ruetschi, D. Jaccard, M. Gabay, D. A. Muller, J. M. Triscone and J. Mannhart, *Science*, 2007, **317**, 1196-1199.
14. J. W. Freeland, K. E. Gray, L. Ozyuzer, P. Berghuis, E. Badica, J. Kavich, H. Zheng and J. F. Mitchell, *Nat Mater*, 2005, **4**, 62-67.
15. J. W. Freeland, J. J. Kavich, K. E. Gray, L. Ozyuzer, H. Zheng, J. F. Mitchell, M. P. Warusawithana, P. Ryan, X. Zhai, R. H. Kodama and J. N. Eckstein, *J. Phys.-Condes. Matter*, 2007, **19**.
16. A. Verna, B. A. Davidson, Y. Szeto, A. Y. Petrov, A. Mirone, A. Giglia, N. Mahne and S. Nannarone, *J Magn Magn Mater*, 2010, **322**, 1212-1216.
17. J. Z. Sun, D. W. Abraham, R. A. Rao and C. B. Eom, *Appl. Phys. Lett.*, 1999, **74**, 3017-3019.
18. H. Yamada, Y. Ogawa, Y. Ishii, H. Sato, M. Kawasaki, H. Akoh and Y. Tokura, *Science*, 2004, **305**, 646-648.
19. M. B. Lepetit, B. Mercey and C. Simon, *Phys. Rev. Lett.*, 2012, **108**.
20. R. Nourafkan, M. Capone and N. Nafari, *Phys. Rev. B*, 2009, **80**.
21. R. Nourafkan and F. Marsiglio, *Phys. Rev. B*, 2011, **84**.
22. E. J. Monkman, C. Adamo, J. A. Mundy, D. E. Shai, J. W. Harter, D. W. Shen, B. Burganov, D. A. Muller, D. G. Schlom and K. M. Shen, *Nat Mater*, 2012, **11**, 855-859.
23. D. Gillaspie, J. X. Ma, H. Y. Zhai, T. Z. Ward, H. M. Christen, E. W. Plummer and J. Shen, *J. Appl. Phys.*, 2006, **99**, 3.
24. J. D. Sayre, K. T. Delaney and N. Spaldin, *arXiv:1202.1431v1 [cond-mat.mtrl-sci]*, 2013.
25. S. Takata, R. Tanaka, A. Hachiya and Y. Matsumoto, *J. Appl. Phys.*, 2011, **110**.
26. K. Fuchigami, Z. Gai, T. Z. Ward, L. F. Yin, P. C. Snijders, E. W. Plummer and J. Shen, *Phys. Rev. Lett.*, 2009, **102**, 066104.
27. J. X. Ma, D. T. Gillaspie, E. W. Plummer and J. Shen, *Phys. Rev. Lett.*, 2005, **95**, 237210.
28. H. Kumigashira, K. Horiba, H. Ohguchi, K. Ono, M. Oshima, N. Nakagawa, M. Lippmaa, M. Kawasaki and H. Koinuma, *Appl. Phys. Lett.*, 2003, **82**, 3430-3432.
29. J. A. Collado, C. Frontera, J. L. Garcia-Munoz, C. Ritter, M. Brunelli and M. A. G. Aranda, *Chem. Mat.*, 2003, **15**, 167-174.
30. J. Sakai, N. Ito and S. Imai, *J. Appl. Phys.*, 2006, **99**.
31. A. K. Heilman, Y. Y. Xue, Y. Y. Sun, R. L. Meng, Y. S. Wang, B. Lorenz, C. W. Chu, J. P. Franck and W. M. Chen, *Phys. Rev. B*, 2000, **61**, 8950-8954.
32. T. Z. Ward, J. D. Budai, Z. Gai, J. Z. Tischler, L. F. Yin and J. Shen, *Nat. Phys.*, 2009, **5**, 885-888.
33. M. Uehara, S. Mori, C. H. Chen and S. W. Cheong, *Nature*, 1999, **399**, 560-563.
34. H. Y. Zhai, J. X. Ma, D. T. Gillaspie, X. G. Zhang, T. Z. Ward, E. W. Plummer and J. Shen, *Phys. Rev. Lett.*, 2006, **97**, 167201.
35. A. J. Millis, *Nature*, 1998, **392**, 147-150.
36. J. Mitra, M. Paranjape, A. K. Raychaudhuri, N. D. Mathur and M. G. Blamire, *Phys. Rev. B*, 2005, **71**, 094426.
37. U. R. Singh, A. K. Gupta, G. Sheet, V. Chandrasekhar, H. W. Jang and C. B. Eom, *Appl. Phys. Lett.*, 2008, **93**, 212503.
38. U. R. Singh, S. Chaudhuri, R. C. Budhani and A. K. Gupta, *J. Phys.-Condes. Matter*, 2009, **21**, 355001.
39. U. R. Singh, S. Chaudhuri, S. K. Choudhary, R. C. Budhani and A. K. Gupta, *Phys. Rev. B*, 2008, **77**, 014404.
40. Z. Gai, S. V. Kalinin, A. P. Li, J. Shen and A. Baddorf, *Advanced Functional Materials*, 2013, 10.1002/adfm.201203425.
41. A. Biswas, S. Elizabeth, A. K. Raychaudhuri and H. L. Bhat, *Phys. Rev. B*, 1999, **59**, 5368-5376.
42. D. V. Evtushinsky, D. S. Inosov, G. Urbanik, V. B. Zabolotnyy, R. Schuster, P. Sass, T. Hanke, C. Hess, B. Buchner, R. Follath, P. Reutler, A. Revcolevschi, A. A. Kordyuk and S. V. Borisenko, *Phys. Rev. Lett.*, 2010, **105**.
43. S. Seiro, Y. Fasano, I. Maggio-Aprile, E. Koller, O. Kuffer and O. Fischer, *Phys. Rev. B*, 2008, **77**.
44. S. J. Liu, J. Y. Juang, J. Y. Lin, K. H. Wu, T. M. Uen and Y. S. Gou, *J. Appl. Phys.*, 2008, **103**.
45. S. Aggarwal and R. Ramesh, *Annu Rev Mater Sci*, 1998, **28**, 463-499.
46. M. J. Calderon, L. Brey and F. Guinea, *Phys. Rev. B*, 1999, **60**, 6698-6704.
47. D. Pesquera, G. Herranz, A. Barla, E. Pellegrin, F. Bondino, E. Magnano, F. Sánchez and J. Fontcuberta, *Nat Commun*, 2012, **3**, 1189.
48. U. Aschauer, R. Pfenniger, S. S. M., T. Grande and N. A. Spaldin, *arXiv:1303.4749v1 [cond-mat.mtrl-sci]*, 2013.
49. C. W. Lin, C. Mitra and A. A. Demkov, *Phys. Rev. B*, 2012, **86**.
50. S. Dong, S. Yunoki, X. T. Zhang, C. Sen, J. M. Liu and E. Dagotto, *Phys. Rev. B*, 2010, **82**, 035118.